\documentclass[conference]{IEEEtran}
\IEEEoverridecommandlockouts

\usepackage[sorting=none]{biblatex}

\usepackage{amsmath,amssymb,amsfonts}
\usepackage{algorithmic}

\usepackage{graphicx}

\usepackage{caption}
\usepackage{subcaption}

\usepackage{textcomp}
\usepackage{xcolor}
\def\BibTeX{{\rm B\kern-.05em{\sc i\kern-.025em b}\kern-.08em
    T\kern-.1667em\lower.7ex\hbox{E}\kern-.125emX}}

\usepackage[font=small]{caption} 

\title{$\mu$TAS: Design and implementation of Time Aware Shaper on SmartNICs to achieve bounded latency}

\author{
    \IEEEauthorblockN{Joydeep Pal, Deepak Choudhary, Nithish Krishnabharathi Gnani, Chandramani Singh, and T. V. Prabhakar}
    \IEEEauthorblockA{\textit{Department of Electronic Systems Engineering} \\
    \textit{Indian Institute of Science
     Bengaluru, India}\\
    joydeeppal@iisc.ac.in, deepakcl@iisc.ac.in, nithishgnani@iisc.ac.in, chandra@iisc.ac.in, tvprabs@iisc.ac.in}
}
\date{20-April-2023}

\begin{document}

\renewcommand{\baselinestretch}{0.98}\selectfont

\maketitle

\begin{abstract}
Time-Aware Shaper (TAS) is a time-triggered scheduling mechanism that ensures bounded latency for time-critical Scheduled Traffic (ST) flows. The Linux kernel implementation (a.k.a TAPRIO) has limited capabilities due to varying CPU workloads and thus does not offer tight latency bound for the ST flows.
Also, currently only higher cycle times are possible. Other software implementations are limited to simulation studies without physical implementation.
In this paper, we present $\mu$TAS, a MicroC-based hardware implementation of TAS onto a programmable SmartNIC. $\mu$TAS takes advantage of the parallel-processing architecture of the SmartNIC to configure the scheduling behaviour of its queues at runtime. To demonstrate the effectiveness of $\mu$TAS, we built a Time-Sensitive Networking (TSN) testbed from scratch. This consists of multiple end-hosts capable of generating ST and Best Effort (BE) flows and TSN switches equipped with SmartNICs running $\mu$TAS. Time synchronization is maintained between the switches and hosts. Our experiments demonstrate that the ST flows experience a bounded latency of the order of tens of microseconds.
\end{abstract}


\section{Introduction} \label{sec:intro}

Tactile Internet applications such as telesurgery require
haptic feedback with a round-trip time of a few milliseconds. 
In telesurgery, the doctors perform robotic surgery remotely, communicating over the Internet, and the associated tactile flows warrant stringent latency bounds of the order of milliseconds\cite{3gpp}.
Existing best-effort 802.1p Ethernet based networks, with priority scheduling for Virtual Local-Area Network (VLAN) traffic fail to provide hard latency guarantees for tactile flows. This is mainly because of the queuing delay of the flows competing for the same outgoing port on a switch.
We leverage the concept of Time-Sensitive Networking (TSN) which promises bounded low-latency  using time-slotted data transmission~\cite{nasrallah2018ultra}. TSN is a set of standards being developed by the TSN Task Group, a part of the IEEE 802.1 Working Group. Furthermore, IEEE 802.1Qbv proposes Time-Aware Shaper~(TAS) for Ethernet switches to minimize queueing latency and to guarantee latency bounds.



To develop novel networking paradigms and to evaluate real-world performance, a range of programmable network hardware have been developed for rapid packet processing and prototyping the desired behaviour. 
Smart Network Interface Cards (SmartNICs) are one category of such networking hardware. To program SmartNICs, network-domain languages are being developed with P4~\cite{P4_user_guide} being the most notable one.  

In this work, we build a TSN switch from scratch using SmartNICs and demonstrate the implementation challenges for $\mu$TAS, our time aware traffic shaper using P4 and MicroC. 

This paper is structured as follows:  Section \ref{sec:background} reviews the standards relevant to this work; Section \ref{sec:related-work} compares $\mu$TAS with related work; Section \ref{sec:system-design} presents detailed implementation of $\mu$TAS; Section \ref{sec:evaluation} evaluates and demonstrates the effectiveness of $\mu$TAS and discusses the results; and Section \ref{sec:conclusion} summarises the work and points to a few future directions.

Following are our contributions:
\begin{itemize}
    \item We build a TSN testbed and demonstrate $\mu$TAS in a network comprising of two switches
    \item We present the first open source programmable TSN hardware solution using SmartNICs
    \item We implement support for PTP based hardware synchronisation on SmartNICs using additional hardware. Our PTP synchronisation messages are carried in in-band channels. 
    
\end{itemize}

\section{Background and Motivation} \label{sec:background}


\subsection{Time-Sensitive Networking}
TSN is a set of Layer-2 Ethernet enhancements defined by the IEEE 802.1 TSN Task Group to enable deterministic communication over switched Ethernet networks. TSN offers bounded latency to the time-critical flows, also called \textbf{Scheduled Traffic (ST)} flows, even in the presence of \textbf{Best Effort (BE)} traffic which is non-critical. To ensure bounded latency for the ST flows, it defines 802.1Qbv TAS  which works in conjunction with 802.1AS Timing and Synchronisation. Thus, TSN switch implementation requires the following components at the least.
\begin{enumerate}
    \item TAS implementation: TSN needs a time-slotted scheduling mechanism for Ethernet networks. Time is divided into slots and each data communication is restricted to one slot. It is however possible for a flow to request  multiple time slots. The notion of Cycle Time (CT) is assigned to a fixed number of time slots, which is repeated over time. Slot-based communication has the advantage of handling multiple concurrent ST and BE flows. The Ethernet switch controller implements a Gate Control List (GCL) for each egress port on the switch to ensure that packets pertaining to a flow are dequeued at its dedicated time slot. 
    \item Time Synchronization: One of the core objectives of TSN is to ensure that all the devices in a TSN Network are time synchronized. TSN uses generic Precision Time Protocol (gPTP) as defined in the IEEE 802.1AS standard~\cite{ieee8021as} for time synchronization, which is based on IEEE 1588 (PTP)~\cite{ieee1588} standard. Time in TSN networks is usually distributed from the system with the most accurate time source, chosen by the Best Master Clock Algorithm (BMCA), directly through the network. PTP utilizes Ethernet frames to distribute time synchronization information.
\end{enumerate}



\subsection{Programmable data planes and TSN}
Programmable data planes allow the creation of custom forwarding plane behaviors to meet the specific requirements of TSN. This allows flexible implementation of custom scheduling mechanisms such as TAS. With programmable hardware, there is more control over hardware  operations such as packet parsing, adding a custom header for relaying queueing and timing information at line rate, optimizing flow control, and configuring the traffic manager for meeting the requirements of a real-time TSN Network. Programmable data planes can be easily integrated with SDN controllers, which makes it easier to configure and manage TSN Networks. It is also possible to have more fine-grained control over traffic management, which allows to define different levels of Quality of Service (QoS) for different traffic classes. It is useful for TSN as it  enables us to give priority to ST flows over the BE flows.

\section{Related Work} \label{sec:related-work}


Scheduling algorithms for TSN have been studied and prototyped in hardware in \cite{sivaraman2016programmable},\cite{alcoz2020sp}. 
Sivaraman et al.\cite{sivaraman2016programmable} present a design for programmable packet schedulers for a network switch without hardware redesign. Alcoz et al.\cite{alcoz2020sp} also design a programmable packet scheduler using priority queues that can work at line rate, on Intel Tofino switches. The authors compare their scheduling to FIFO scheduling using multiple priority queues in terms of throughput and flow completion times for traffic flows. They achieve flow completion times of ~1 ms. However, they do not consider time critical flows. 
These studies do not provide any studies on its effect on latency performance for ST and BE flows with multiple TSN switches built on hardware. 

Below we present existing approaches and their limitations with regard to achieving deterministic latency.

\textbf{Software-based approach}
is a way to evaluate the performance of TAS by utilising the services of Time Aware Priority Shaper (TAPRIO\cite{taprio}), a Linux based queueing discipline module, which implements TAS and demonstrates isolation of traffic flows. Kumar et al. \cite{kumar2020coupling} applied TAPRIO on Mininet based emulated switches along with source routing. The authors demonstrate an end-to-end latency in a network topology comprising of a source host, two TAPRIO-enabled switches and a destination host. A best case end-to-end latency of $100 \mu s$ was observed. This delay is mostly due to  kernel and user space processing time. Also, several underlying assumptions such as propagation delay and packet processing delays are ignored. If a software-based approach is deployed in practice, the delays might vary significantly and make this unsuitable for TSN applications. Ulbricht et al. \cite{ulbricht2021emulation} also used Mininet-emulated switches with TAPRIO. They demonstrate that the processing delay in such a switch is $2 \mu s$. The implementation of the same on TrustNode hardware switches demonstrates processing latency of $1.4 \mu s$. Falk et al. \cite{falk2019nesting} present a simulation on TAS by extending a network simulator called OMNET++. They demonstrate a worst-case delay of $32 \mu s$. Such software-based approaches suffer from variable execution times due to its dependence on the OS kernel.

\textbf{Hardware-based approach}
is another way to demonstrate TSN capabilities is by using an off-the-shelf TSN switch or by designing a TSN switch on network hardware.
\cite{grigorjew2021distributed} proposes to do away with TSN by using a reservation protocol executed in the control plane with packet forwarding carried out by a NEC PF5420 switch. They demonstrate bounded round-trip latency of 1 ms. However, they demonstrate with a single switch. Also there is no notion of CT.
Jiang et al. \cite{jiang2019simulation} implement a TSN Switch using Cisco Industrial Ethernet (IE 4K) switches. However, it is not possible to set custom CT and GCL. Also, the bound on latency is hard-coded to 1 ms.

Programmable SmartNICs are emerging as a popular choice to design and implement an increasing number of network functions completely on hardware for data-centre networks with vendors like Netronome, Intel, NVIDIA, and others. It provides flexible packet processing at line-rate and high-precision clocks that can support such time-triggered scheduling mechanisms.
SmartNICs have been used to implement scheduling functions in \cite{flowvalve}, \cite{atutxa2021achieving}.
Xi et al.\cite{flowvalve} propose an offload design for the classifying and scheduling functions of the Linux traffic control such as Priority Qdisc (PRIO) and Hierarchy Token Bucket (HTB). However, FlowValve focuses on achieving higher throughput and enforcing network policies at line rate, whereas $\mu$TAS demonstrates end-to-end latency performance and ensures bounded delay. 
Atutxa et al. \cite{atutxa2021achieving} aim to achieve lower latency by processing MQTT packets in data plane using P4. Their work however does not consider hard latency guarantees. 


Implementations of TSN standards are either in a simulated environment or on non-programmable hardware. Our work aims to bridge the gap between programmable data planes and TSN implementations. We leverage the features of programmable SmartNICs, such as flexible packet processing, precise clocks and a configurable traffic manager to program custom scheduling mechanisms using P4 and MicroC programming languages, to design and implement \textbf{P4} - \textbf{T}ime \textbf{A}ware \textbf{S}haper ($\mu$TAS) which resides in the data plane.

\section{System Design} \label{sec:system-design}
 
The $\mu$TAS is a module that runs on the SmartNIC. In our work, we develop a network testbed with the SmartNIC being programmed as a TSN switch, a key building block to achieve our goal.
    We introduce our SmartNIC and describe the hardware features of this networking platform. We then describe our $\mu$TAS implementation including Cycle Time (CT) and Gate Control List (GCL) design.

\subsection{Programmable SmartNIC} \label{sec:system-design:smartnic}
\begin{figure}[htbp]
    \centerline{
    \includegraphics[width=0.45\textwidth]{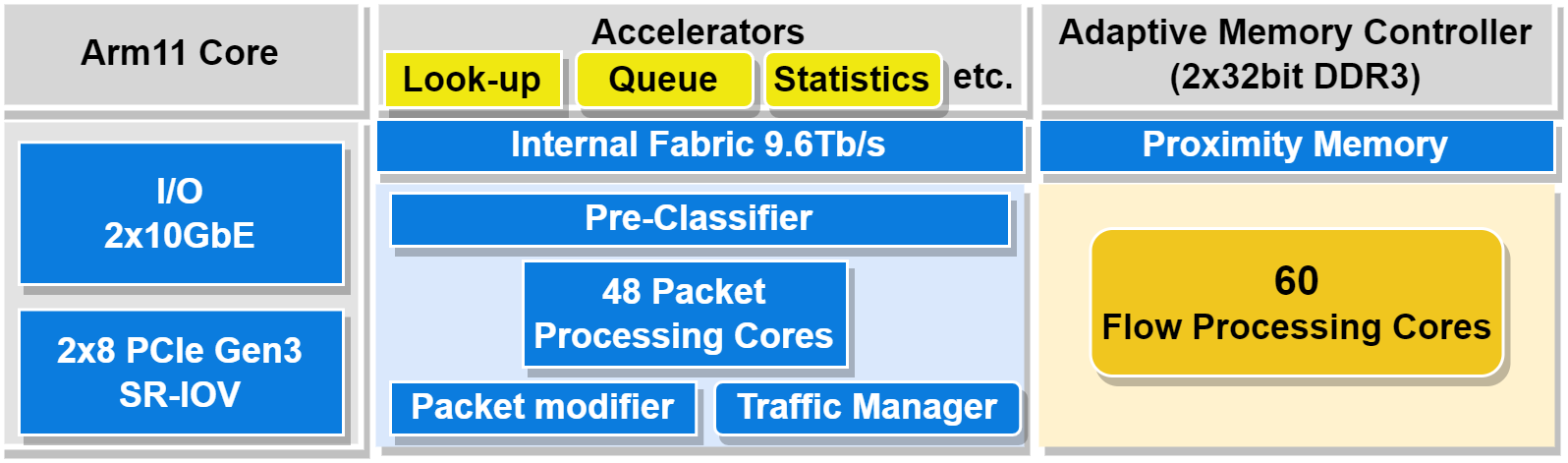}
    }
    \caption{Netronome SmartNIC Architecture \cite{nfp-4000-diag}. Blocks of interest: (a) Flow Processing Cores, (b) Traffic Manager, (c) Queue, (d) Statistics}
    \label{nfp}
\end{figure}

We chose Netronome Agilio CX 2x10GbE for its ability to support P4 programming. Fig. \ref{nfp} represents the architecture and range of features including our blocks of interest. The 60 programmable parallel flow processing cores called MicroEngines (ME) support wire-speed packet processing. It has a single built-in clock source for precise time-keeping with a resolution of one nanosecond. This clock is a single counter which starts from zero each time the system is restarted. Each ME can be programmed independently and has access to the global time  from the SmartNIC's time-keeping registers. Additionally, the Traffic Manager can be configured to handle multiple queues, schedulers and shapers. Specifically, we obtain current timestamps, set scheduling parameters, assign specific queues to a particular flow and obtain queue occupancy levels at runtime. The SmartNIC has a rich set of Control and Status Registers (CSR) that provide configuration control and status updates. These registers can be accessed using the Netronome command line interface (CLI) from the host system (via the nfp-reg commands) or from the data plane using MicroC language.

\subsection{$\mu$TAS}
    Our novel $\mu$TAS implementation supplements other standard Ethernet switch functional modules that are readily available in the Netronome repository. Some examples include bridge, VLAN support, full duplex operation, auto-negotiation etc. $\mu$TAS utilizes the services of several MEs to implement various functions, including forwarding logic and capturing statistics for each flow. We dedicate one ME to implement the packet scheduling function.

While the packet processing pipeline is developed in P4, the scheduling function is implemented in MicroC and integrated with P4 as an extern function. Each packet received at the ingress of the switch port is parsed using P4 to extract the Ethernet, VLAN and IP headers. The VLAN header's ID is used to determine the flow, and a MicroC extern function directs the packet to the flow's corresponding egress queue. 


%

\begin{figure}[htbp]
    \centerline{
    \includegraphics[width=0.42\textwidth]{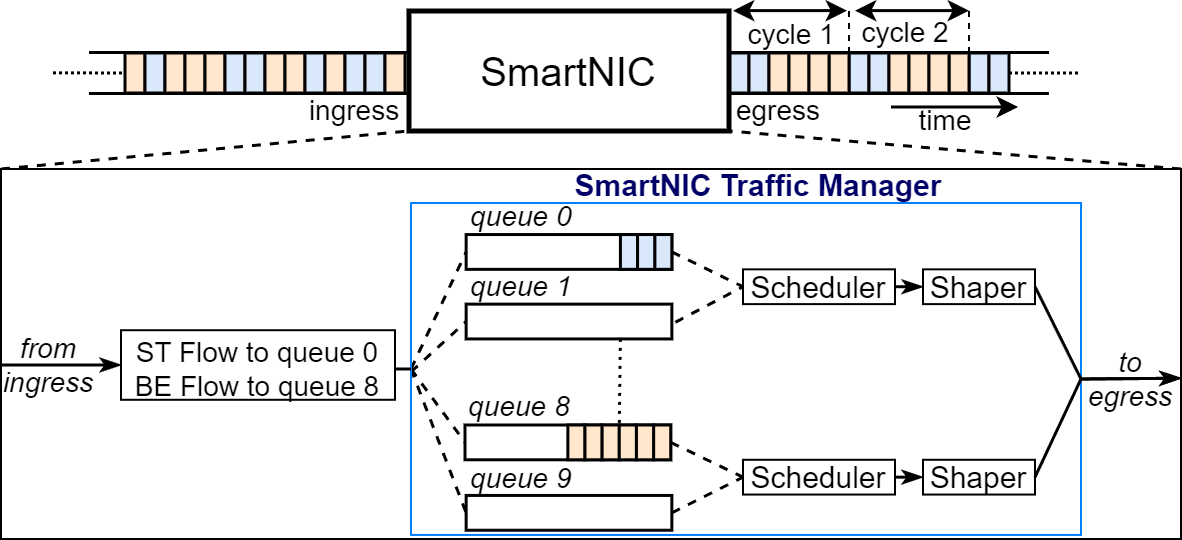}
    }
    \caption{Time-Aware Shaper on SmartNIC}
    \label{fig:system-design:tas}
\end{figure}

\subsubsection{GCL implementation}
 
GCL facilitates the flow of packets in TAS that have bounded latency.
As shown in Fig. \ref{fig:system-design:tas}, our $\mu$TAS implementation ensures temporal isolation of flows at the egress of a port. 
The incoming data is assigned to a queue based on its flow using VLAN ID. While the Scheduler maintains the service time of a queue, the Shaper at the beginning of the queue service time maximises its bandwidth. The GCL, which consists of the tabular entries for opening and closing time of queue gates, is enforced by configuring the Traffic Manager. We assume that GCL is made available by programming SMT solvers such as Z3, which consider traffic flows and their requirements as input.  

\subsubsection{CT implementation}
 
\begin{figure}[htbp]
    \centering
    \includegraphics[width=0.38\textwidth]{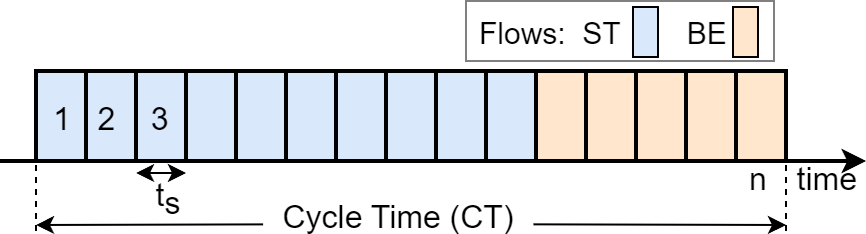}
    \caption{ CT used in TAS}
    \label{fig:system-design:ct}
\end{figure}

Fig. \ref{fig:system-design:ct} describes the CT where flows are segregated in time by allotting time slots. Time is divided into $n$ slots where $t_{s}$ is the length of each slot. The CT repeats itself after $n$ slots. In each CT, flows are allowed to dequeue from the egress queue only in their specific time slot. The standard recommends that in a CT, ST flows are assigned a maximum of 70\% of the time slots. In our implementation, we chose a CT of 5 ms. For simplicity, we considered a single ST flow occupying 70\% of the slots and a single BE flow occupying the remaining 30\%.

The $\mu$TAS's CT implementation is accomplished with a separate MicroC extern function by using the time function available in the SmartNIC. To make this available to the scheduling function, the control plane programs the set of data plane memory registers. The scheduler is assigned the responsibility to map time slots to queues. For E.g., from Fig. \ref{fig:system-design:tas}, ST flow from queue 0 is assigned slot 1 shown in Fig. \ref{fig:system-design:ct}. Since $\mu$TAS requires the beginning of the time slot to schedule a traffic flow, it probes the time CSR for the current time ($t_{curr}$). 

\begin{equation} \label{equation:ct}
t_{debug} = (t_{curr})mod(CT)
\end{equation}

$t_{debug}$, which is the modulus of $t_{curr}$ with CT, is calculated as shown in the following equation \ref{equation:ct}. If $t_{debug}$ is equal to the start of a time slot, $\mu$TAS makes a call to the Traffic Manager to configure its shapers appropriately. For example, in a CT of 10ms, if ST flow requires queue ($q_{0}$) to be serviced from 0 to 7 ms, and BE flow requires queue ($q_{1}$) from 7 to 10 ms, then at $t_{debug}$ = 0 ms, the Traffic Manager maximises the $q_{0}$ bandwidth and restricts the $q_{1}$ bandwidth to zero, and vice-versa for $t_{debug}$ = 7 ms.



\subsection{Time Synchronization} \label{sec:system-design:time-sync}
 
Ethernet switches in TSN require tight synchronisation to enforce GCL on each switch and thus maintain end-to-end latency bound for ST flows. 
To realize this, it is imperative to synchronise the hardware clocks on SmartNICs. Since the timeslot in a CT is in the millisecond range, it is expected that
the synchronisation accuracy after budgeting for overheads should either be in micro or nanoseconds. Protocols such as Network Time Protocol and Precision Time Protocol (PTP) with software timestamping fall short of providing synchronisation with nanosecond accuracy. 

\begin{figure}[htbp]
    \centerline{
    \includegraphics[width=0.40\textwidth]{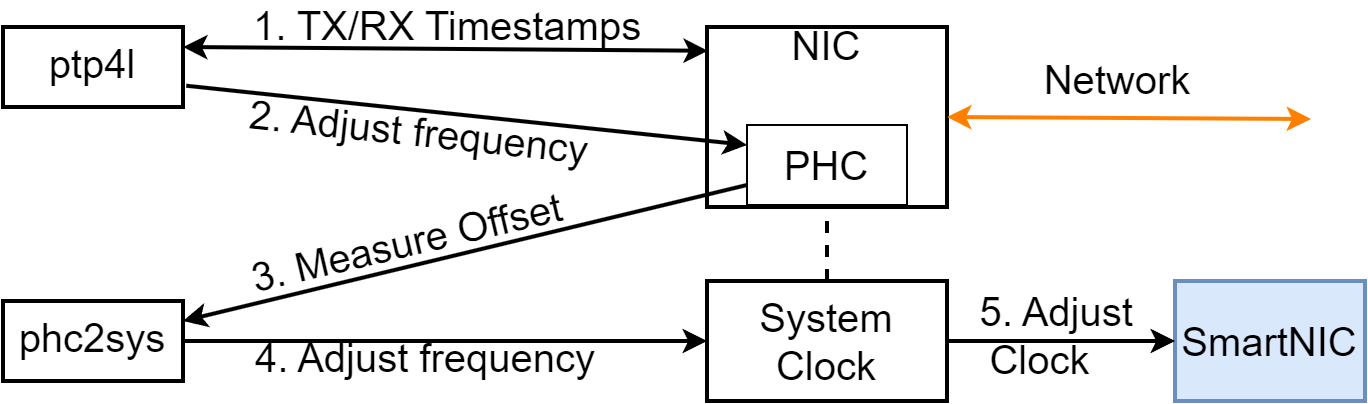}
    }
    \caption{Adjusting SmartNIC clock using linuxptp}
    \label{ptp_me}
\end{figure}

\textbf{Synchronizing the hosts in a network:}
To facilitate hardware synchronisation, an \emph{Intel X520 10-Gigabit dual-port NIC} which supports hardware clock, is installed in each host (refer Fig. \ref{fig:testbed}).
We now describe the procedure to synchronise hosts' NIC hardware clocks, known in the PTP domain terminology as PTP Hardware Clock (PHC). Fig. \ref{ptp_me} shows the message sequence where PHC peers are synchronised using \emph{linuxptp}'s sub-module \emph{ptp4l}. The best PHC is chosen using the BMCA. Another sub-module of linuxptp, \emph{phc2sys} ensures that in a given host, the PHC synchronises the system clock i.e CLOCK\_REALTIME. We obtain a synchronisation error below 10 ns.

\begin{figure}[htbp]
    \centerline{
    \includegraphics[width=0.48\textwidth]{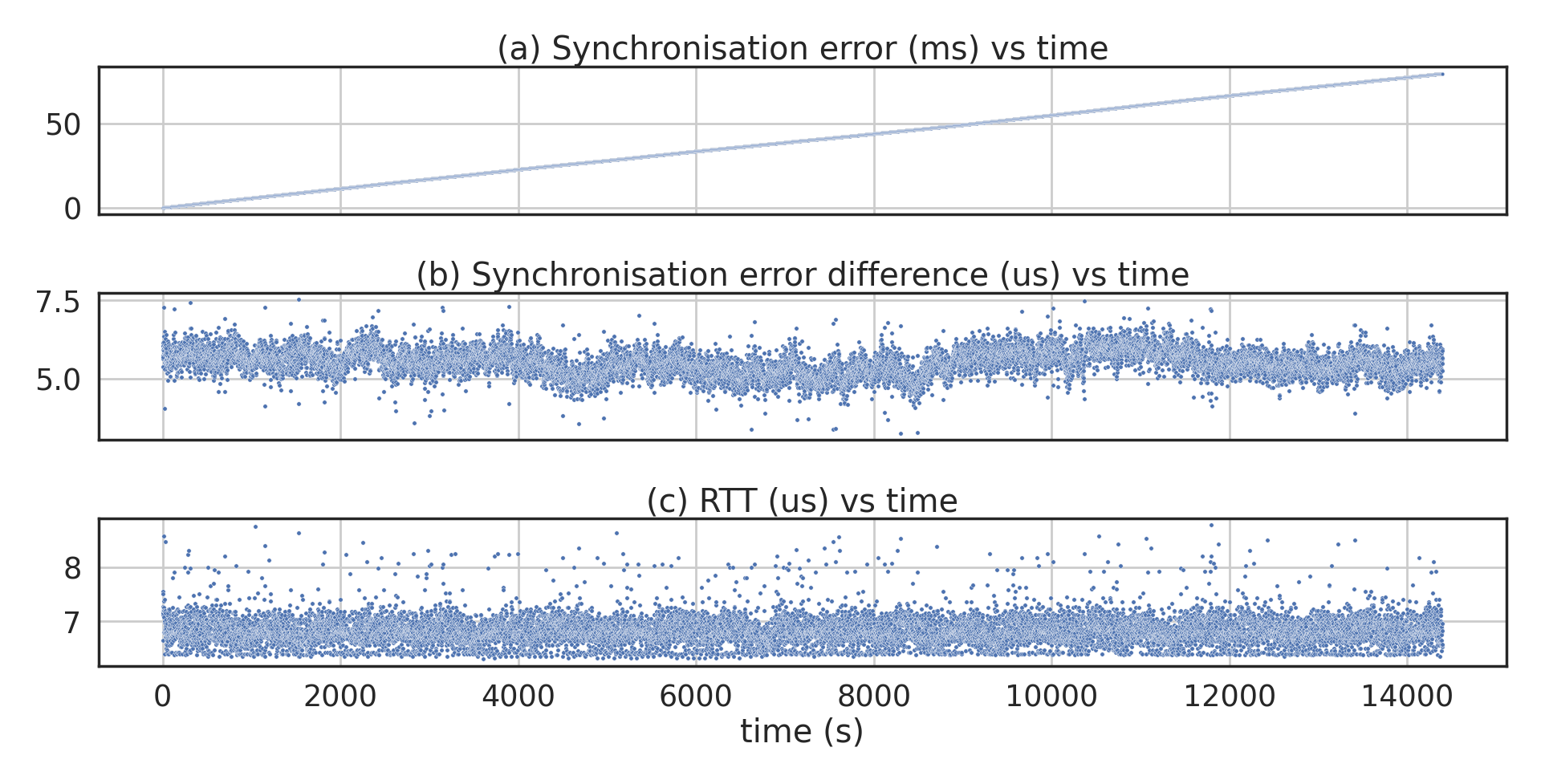}
    }
    \caption{Synchronisation error between SmartNIC clocks}
    \label{fig:system-design:smartnic_clock_sync_error}
\end{figure}

\begin{figure}[htbp]
    \centerline{
    \includegraphics[width=0.48\textwidth]{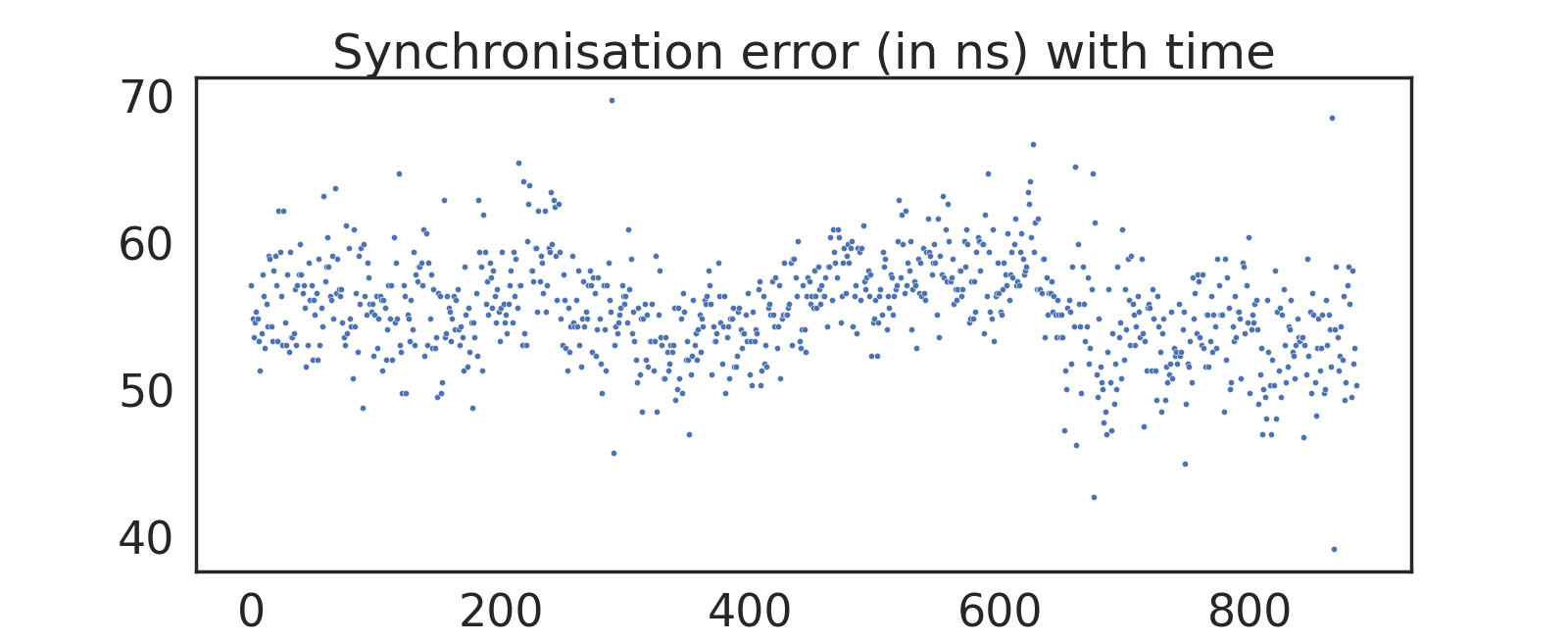}
    }
    \caption{Synchronisation error after configuration phase}
    \label{fig:system-design:smartnic_clock_sync_error_sync}
\end{figure}

\textbf{Synchronizing the SmartNICs:}
While PTP works satisfactorily if the NIC card supports hardware clocks, unfortunately, our SmartNIC does not support this protocol and therefore, our next step is to synchronise the SmartNIC clock to the host's system time which is synchronised in the previous step. We implemented a custom in-band synchronisation scheme. To estimate the clock drift between the two spatially separated SmartNICs, we design an experiment that comprises of two phases.
In the data collection phase, the source host connected to the first SmartNIC sends out a packet containing the ME timestamp. The downstream second SmartNIC in turn appends its own timestamp and returns the packet to the first SmartNIC which again appends its ME timestamp and then forwards the same back to the host. Several hundreds of such packets over multiple days are collected and analyzed for drift to ensure repeatability of measurements. Fig. \ref{fig:system-design:smartnic_clock_sync_error}(c) shows the Round Trip Time (RTT) between the first and second SmartNIC. An RTT of $6$-$9 \mu s$ reflects the propagation delay and processing delay in the SmartNICs. Meanwhile, Fig. \ref{fig:system-design:smartnic_clock_sync_error}(a) shows the synchronisation error due to clock drift. It is observed that the drift varies linearly with time and our calculations indicate that it varies 20 ms over a period of 1 hour, and perhaps one may be able to compensate for the same. The consecutive differences of the previous plot are showed in Fig. \ref{fig:system-design:smartnic_clock_sync_error}(b), i.e we measure the slope for the linear drift.
In the configuration phase, our in-band synchronisation starts from the first SmartNIC where it transmits a packet that contains the ME timestamp. The downstream receiving switch accepts this packet and configures its ME timestamp with this timestamp. Since the measured clock drift between the two switches is linear, a special packet containing the slope and intercept obtained from the data collection phase is supplied to the second switch to perform automatic compensation. Fig. \ref{fig:system-design:smartnic_clock_sync_error_sync} shows that the synchronisation error has been brought down to tens of nanoseconds after configuration phase.

\section{Evaluation} \label{sec:evaluation}
   
We describe our network testbed to evaluate and compare the performance of $\mu$TAS with Round-Robin scheduling, Priority Scheduling and TAPRIO. We show the latency comparison 
for ST and BE flows. Each queue's runtime occupancy level is obtained from the SmartNIC using $\mu$TAS.

\begin{figure}[htbp]
    \centering
    \includegraphics[width=0.48\textwidth]{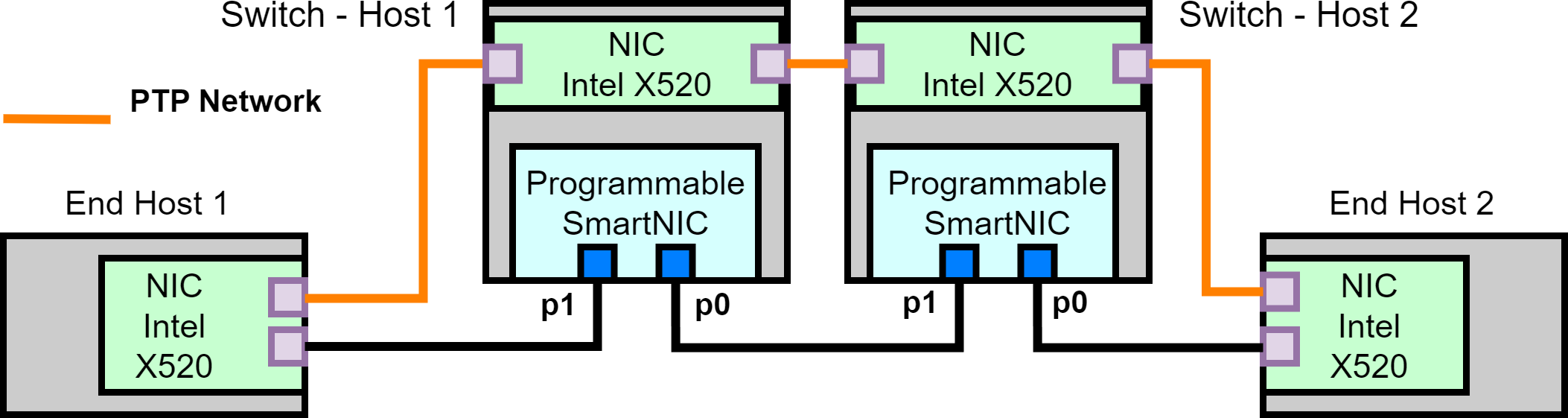}
    \caption{TSN testbed}
    \label{fig:testbed}
\end{figure}

\textbf{Evaluation environment:} Fig. \ref{fig:testbed} shows the testbed used in our experiments. It consists of two TSN switches and two end-hosts connected in series using one metre of full-duplex optical fibre Ethernet links  as the connection medium. The end-hosts are each equipped with a dual-port NIC mentioned previously, supported by an Intel i5 CPU and 8GB of RAM. 
The TSN switches are built using SmartNICs installed on hosts, where each host is equipped with a Asus Z690 motherboard with four PCIe Gen 3x8 slots, a 12-core Intel i5 CPU and 32GB of RAM. All the hosts run Ubuntu OS. The SmartNIC is inserted into one of the PCIe Gen 3x8 slots.
Host-1 and Host-2 are the traffic generator and receiver, respectively. Host-1 transmits ST, and BE flows with a packet transmission rate of 10 Mbps and a packet length of 1000 bytes.

\begin{figure*}[htbp]
    \centering
    \begin{subfigure}[b]{0.4\textwidth}
        \centering
        \includegraphics[width=\textwidth]{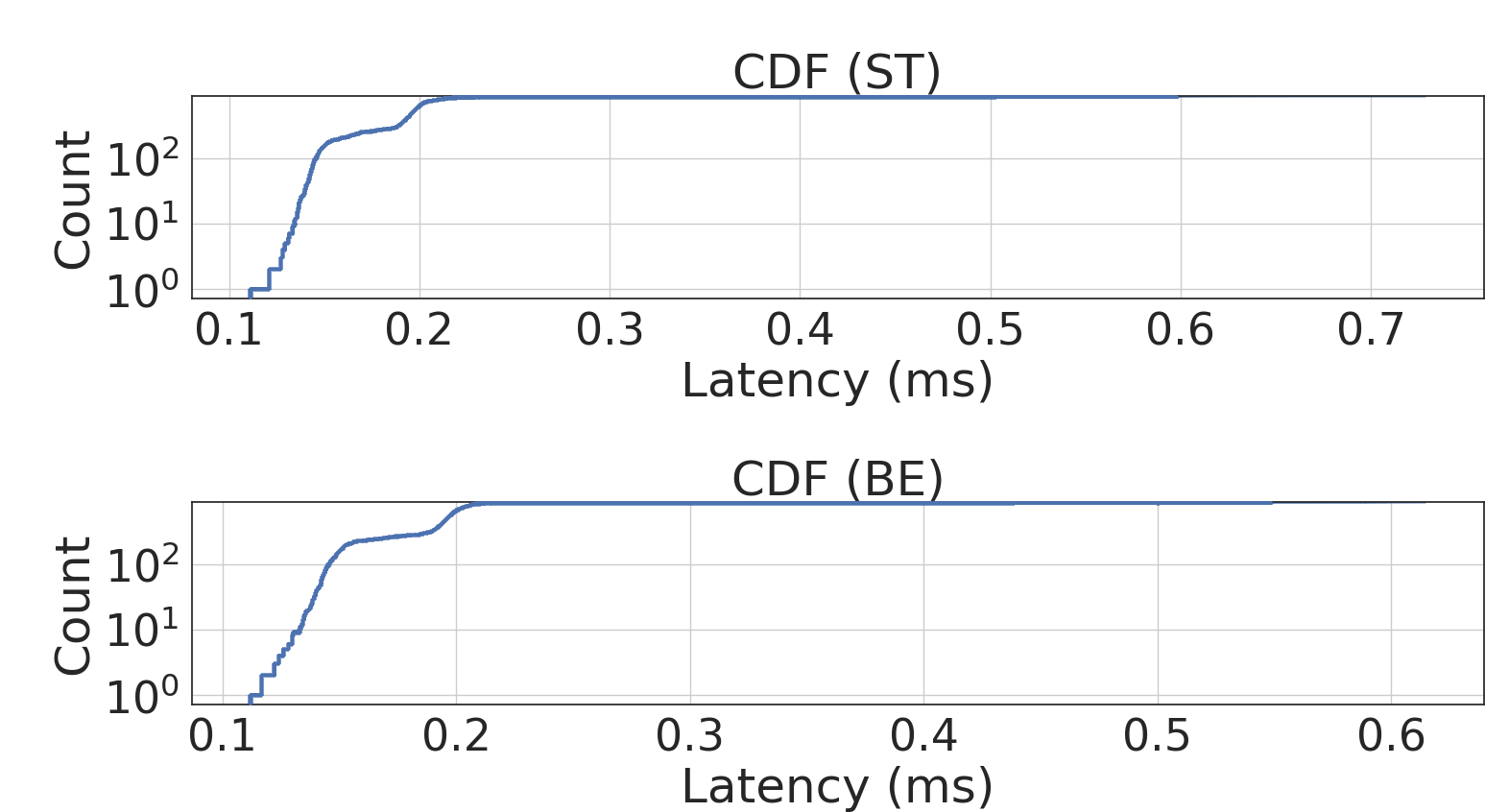}
        \caption{Default Round-Robin scheduling}
        \label{default-sched}
    \end{subfigure}
    \begin{subfigure}{0.4\textwidth}
        \includegraphics[width=\linewidth, height=4cm]{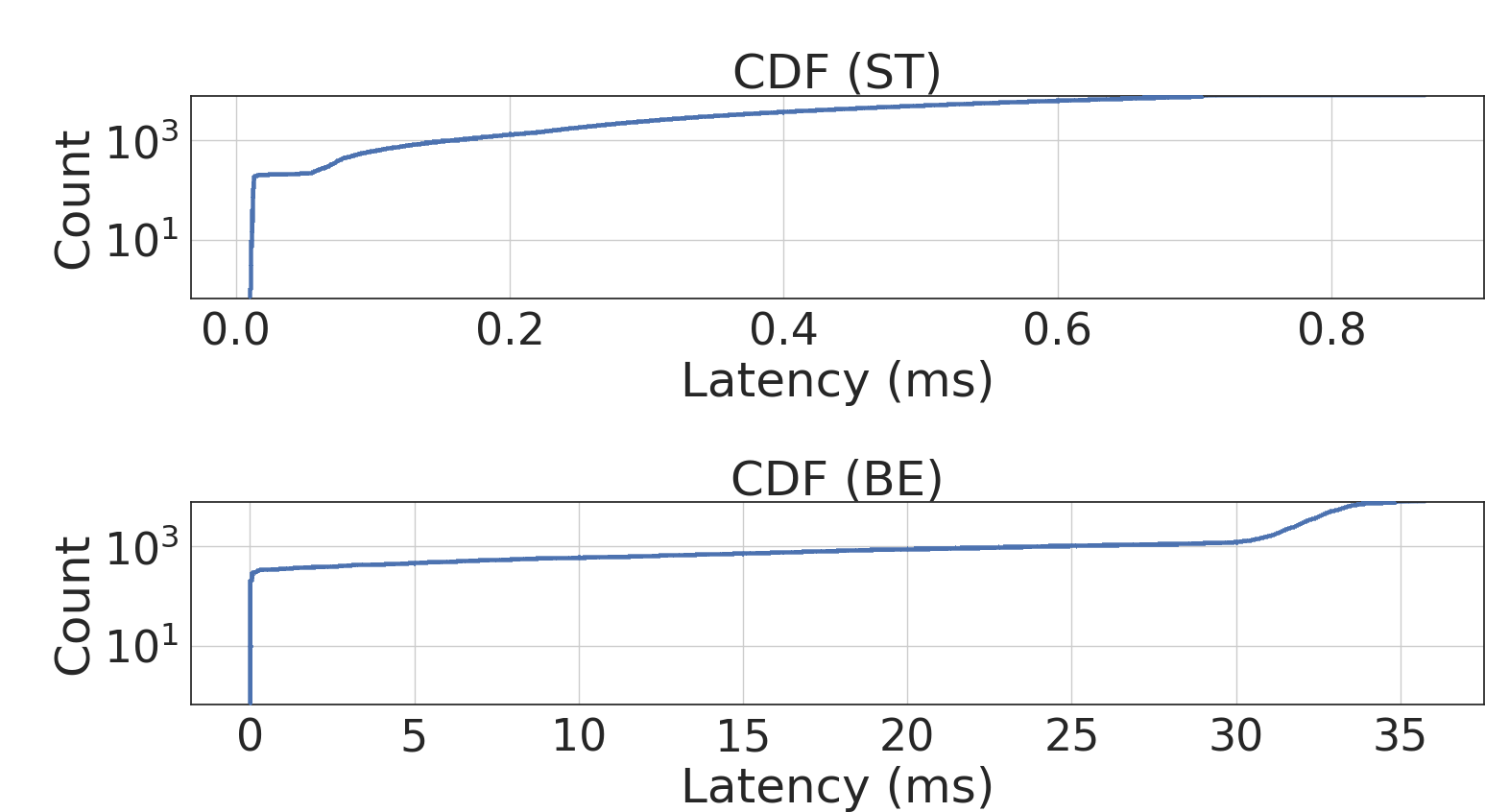}
        \caption{Strict Priority}
        \label{sp-sched}
    \end{subfigure}
    \begin{subfigure}{0.4\textwidth}
        \includegraphics[width=\linewidth, height=4cm]{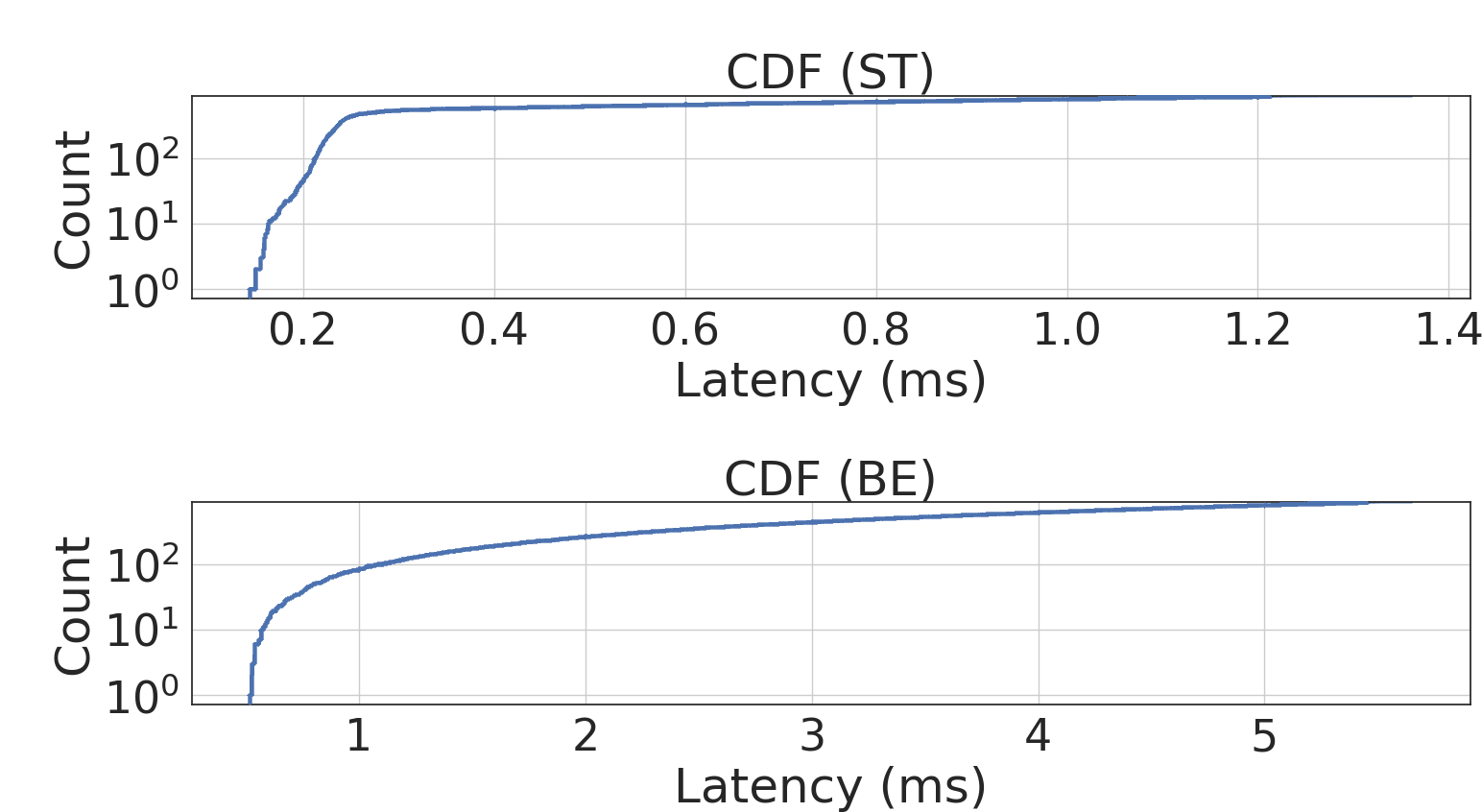}
        \caption{TAPRIO}
        \label{taprio-sched}
    \end{subfigure}
    \begin{subfigure}{0.4\textwidth}
        \includegraphics[width=\linewidth, height=4cm]{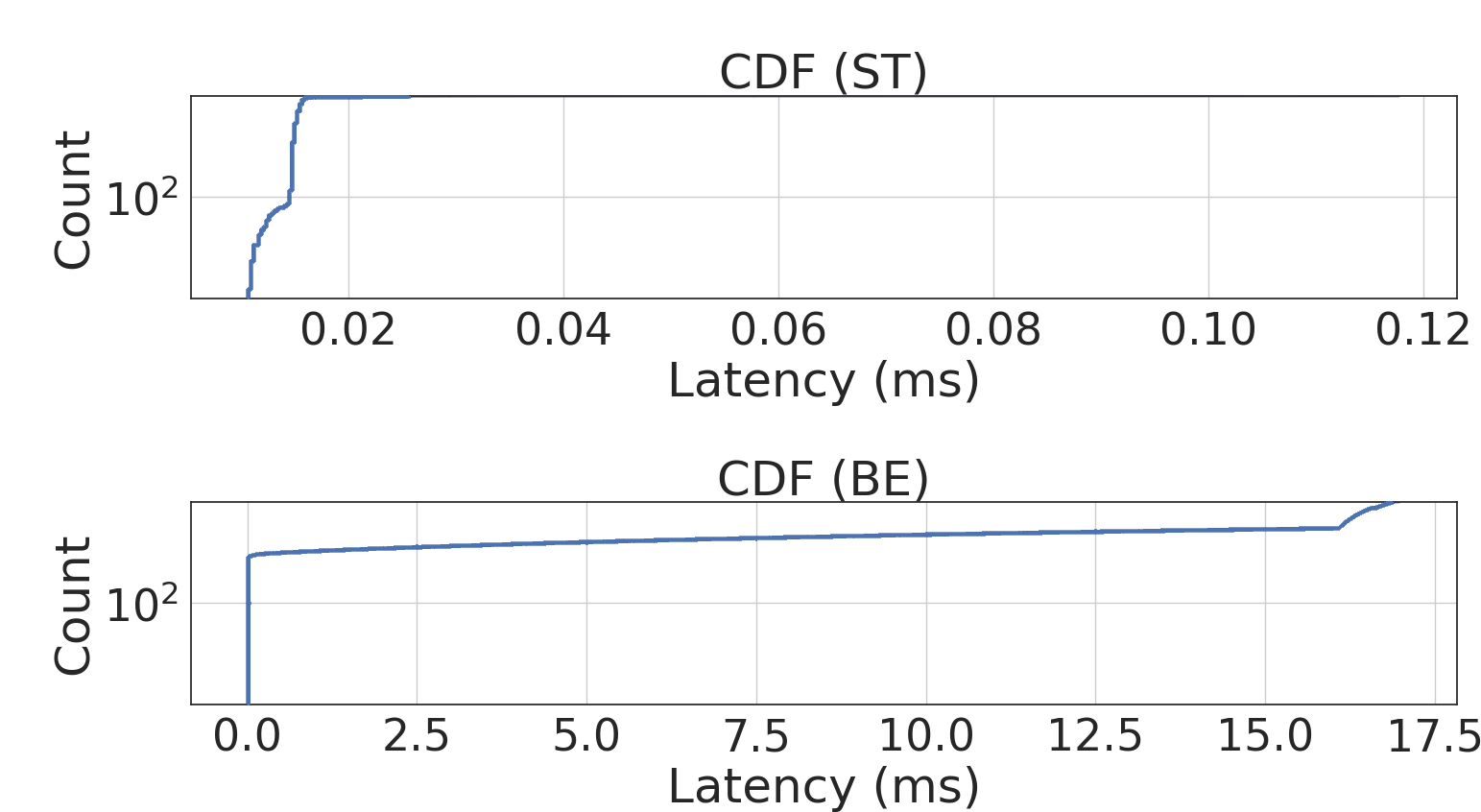}
        \caption{$\mu$TAS}
        \label{P4TAS-sched}
    \end{subfigure}
    \caption{Latency experienced by each packet for (a) Default Round-Robin scheduling, (b) Strict Priority, (c) TAPRIO and (d) $\mu$TAS}
    \label{fig:latency}
\end{figure*}

\textbf{Performance baseline:} Since the goal of our work is to ensure a bounded latency for ST flows in the presence of BE flows, we compare the latency offered by the following four scheduling algorithms. We first measure the performance of vanilla SmartNIC with the default nic-firmware installed. It is an official firmware designed by Netronome to be used as a performance benchmark for custom implementations.
We compare this default scheduling behaviour with Strict Priority (SP), TAPRIO and $\mu$TAS, all running on Netronome, with SP and $\mu$TAS being a complete data plane implementation on hardware.
\begin{itemize}
    \item Case (a) - Switch is loaded with nic\_firmware and uses default scheduling, i.e. Round-Robin scheduling
    \item Case (b) - Switch is configured for SP scheduling. ST flows are assigned higher priority over BE flows at the source.
    \item Case (c) - Switch is configured as a TSN switch by loading it with nic\_firmware and then applying the Linux TAPRIO module at the egress ports. 
    \item Case (d) - Switch is configured as a TSN switch with $\mu$TAS scheduling by loading our P4-based program on the SmartNIC
\end{itemize}

\subsection{Latency experienced by ST and BE flows}
We run a utility called \textit{tshark} in the end-hosts to capture packets and store them in a Packet Capture (pcap) file. These files are analyzed to obtain the latency experienced by each packet, where packets are uniquely identified using a combination of VLAN and IP headers. Since the hosts are time-synchronized as described in \ref{sec:system-design:time-sync}, latency measurements can be obtained by a direct comparison of the resulting pcap data files timestamped at the transmitter and receiver.
We analyze the data by extracting the packet timestamps, VLAN IDs, and IP sequence number. Note that the egress port bandwidth is restricted to a value less than the combined transmitting bandwidth, which creates congestion at the outgoing links of the switch and leads to queueing of packets and this reflects in the end-to-end latency.  
Fig. \ref{fig:latency} demonstrates the latency for each packet in the above-mentioned cases (a)-(d). Initially, the first few packets experience the lowest latency and as more packets arrive at the switch, queueing takes place at the egress ports, and this is reflected in the increase in latency. We present the cumulative density function (CDF) of end-to-end latency for both flows between the end hosts. A CDF helps in easily observing any breaches over the latency bound. In the case of (a), the default round-robin scheduling, we observe that both flows experience similar latencies in the range of 0.1-0.6 ms. In the case of (b) where SP is enabled in the SmartNIC, packets of ST flow experience lower latency compared to BE, although we also observe that a significant number of packets suffer from latencies in the range of 0.1-0.7 ms, and a clear bound cannot be established. We set a CT of 5 ms. Once we apply TAPRIO, we observe that although ST flows experience lower latency than BE in this case, the latency bound is breached for ST flow. This can be attributed to variability in kernel service times.
For the same flows in case (d), we observe that a bounded latency of 0.02 ms, as desired, is achieved for ST flow with $\mu$TAS i.e. a 10 times reduction in end-to-end latency than what can be achieved using SP and TAPRIO.



\subsection{Queue occupancy levels of the two queues in the switch}
   
The SmartNIC processes packets at very high speeds resulting in packet processing times in nanoseconds. To obtain the queue occupancy, we call a function to read the queue statistics and store it in a special block of memory called real-time symbol memory. We read these special memories from the control plane using the \emph{nfp-rtsym} tool in Netronome CLI and log this data periodically to a csv file. Since the control plane is placed in the Linux kernel, the lowest time period with which we can probe this memory is 100 milliseconds. However, there are certain situations where such a high period is insufficient. Consider the case where it may so happen that a queue builds up and is drained in less than 100ms, which may fall between two reads of the queue. This leads to missing several data points for queue occupancy in this time interval.  To overcome this challenge, we develop a MicroC program to run in a separate ME. This program obtains queue occupancy at the smartNIC's clock speed, sums it up using a temporary variable \emph{queue-level-temp} at an interval of 100 ms. At the end of this interval, it averages out the summed variable, writes this value to the special memory and resets the \emph{queue-level-temp} to zero. We probe this value at 100 ms but without missing any data points. We set the queue size to 64 packets.

\begin{figure}[htbp]
    \centering
    \includegraphics[width=0.48\textwidth]{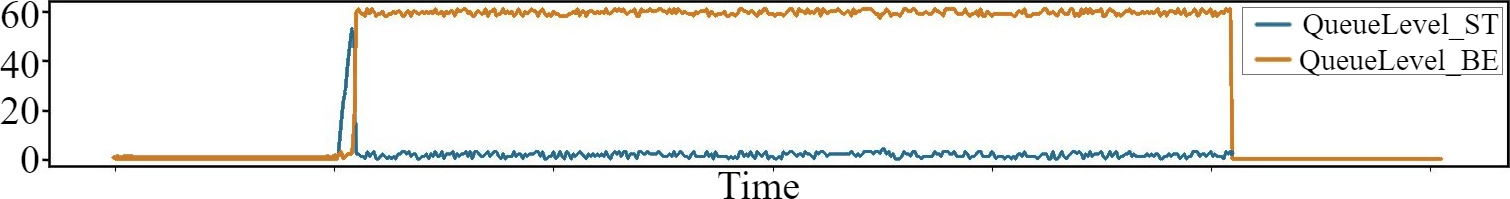}
    \caption{Variation of Queue occupancy levels with time for ST and BE egress queues}
    \label{fig:evaluation:qlevel}
\end{figure}

Fig. \ref{fig:evaluation:qlevel} illustrates the plot of queue occupancy levels for BE and ST egress queues to demonstrate that queue build-up, and hence queueing latency for ST flows, is lower than that for BE flows. We observe that  there is an initial queue build-up for both flows as we start their transmission from the source. Subsequently, the ST queue level remains under 3 while BE queues continuously get exhausted at around 64. As packets egress from the queues, the occupancy levels fall before rising up again due to queueing of new packets arriving at the switch.






\section{Conclusion and Future Work} \label{sec:conclusion}
This paper presents $\mu$TAS, a P4-based system design for TAS on a programmable SmartNIC. $\mu$TAS leverages the multi-core parallel architecture, configurable traffic manager and the accurate clocks of the SmartNICs to synchronize and implement TAS among two switches. We evaluate $\mu$TAS on our TSN hardware testbed and observe that the ST flows have a bounded latency in the presence of BE flows. To the best of our knowledge, ours is the first implementation of TAS on SmartNICs. Our future work explores understanding the SmartNIC architecture in more depth and to use P4 to harness its full capability to enhance our TSN switch and incorporate other IEEE 802.1 standards such as 802.1CB for reliability. We also look at automated configuration of GCL using data-driven approaches by processing and analyzing traffic flows at runtime.


\section*{Acknowledgements} \label{ack}
This work was supported in part by the Ministry of Electronics and Information Technology (MeitY), Government of India and in part by Centre for Networked Intelligence (a Cisco CSR initiative) at Indian Institute of Science, Bangalore.

\printbibliography
\end{document}